\documentclass{pasj00}
\begin{document}
\SetRunningHead{R. Kawabata \& S. Mineshige}{Radiative Spectra in Black Hole X-ray Binaries}
\Received{2009/11/16}%{yyyy/mm/dd}
\Accepted{2010/3/4}%{yyyy/mm/dd}

\title{Radiative Spectra from Disk Corona and Inner Hot Flow in Black Hole X-ray Binaries}

%%% begin:list of authors
% Do NOT capitalize all letters in "textsc".
\author{Ryoji \textsc{Kawabata} and Shin \textsc{Mineshige}} %
\affil{Department of Astronomy, Kyoto University, Kyoto 606-8502}
\email{kawabata@kusastro.kyoto-u.ac.jp}

%%% end:list of authors
%%% Please use the following style in case that sorting by 
%%% affilation is impossible. 
%
% \author{%
%   D-Firstname \textsc{D-Familyname}\altaffilmark{1}
%   E-Firstname \textsc{E-Familyname}\altaffilmark{1,2}
%   and
%   F-Firstname \textsc{F-Familyname}\altaffilmark{2}}
% \altaffiltext{1}{Address of Institute}
% \email{ddddd@xxx.xxx.xx.xx}
% \email{eeeee@xxx.xxx.xx.xx}
% \altaffiltext{2}{Address of Institute}

%% `\KeyWords{}' always has to be placed before `\maketitle'.
\KeyWords{accretion, accretion disks---black hole physics---hydrodynamics---radiation mechanisms: thermal} %Do NOT move this preamble from here!

\maketitle

%%%%%%%%%%%%%%%%%%%%%%%%%%%%%%%%%%%%%%%%%%%%%%%

\begin{abstract}

To understand the origin of hard X-ray emissions from black hole X-ray binaries during their low/hard states, we calculate the X-ray spectra of black-hole accretion flow for the following three configurations of hot and cool media: (a) an inner hot flow and a cool outer disk (inner hot flow model), (b) a cool disk sandwiched by disk coronae (disk corona model), and (c) the combination of those two (hybrid model). 
The basic features we require for successful models are
(i) significant hard X-ray emission whose luminosity exceeds that of soft X-rays,
(ii) high hard X-ray luminosities in the range of $(0.4 - 30)\times 10^{37}$ erg s$^{-1}$,
and (iii) the existence of two power-law components in the hard X-ray band
with the photon indices of $\Gamma_{\rm s}\sim 2 > \Gamma_{\rm h}$,
where $\Gamma_{\rm s}$ and $\Gamma_{\rm h}$ are the photon indices
of the softer ($< 10$ keV) and the harder ($> 10$ keV) power-law components, respectively.
Contribution by non-thermal electrons nor time-dependent evolution are not considered.
We find that Models (a) and (b) can be ruled out, since the spectra are always dominated by the soft component, and since only one power-law component, at most, can be reproduced.
Only Model (c) can account for sufficiently strong hard X-ray emissions, as well as the existence of the two power-law components, for a large ratio of the accretion rate in the corona to that in the thin disk. 
The outer disk corona (where the Compton $y$-parameter is smaller, $y < 1$) produces the softer power-law component with photon index of $\Gamma_{\rm s} \sim 2$, whereas the inner hot flow (where $y \gtrsim 1$) generates the harder component with $\Gamma_{\rm h} < 2$.
This model can also account for the observed relationship between the photon index and the reflection fraction.

\end{abstract}

\section{Introduction}

It is well known that black hole X-ray binaries show two basic spectral states: the high/soft state appearing at high luminosities (typically 10\% $L_{\rm Edd}$, where $L_{\rm Edd}$ is the Eddington luminosity) and the low/hard spectral state at low luminosities, typically a few percent of $L_{\rm Edd}$, by factors of a few smaller than that of the high/soft state (McConnell et al. 2002).
In the high/soft state X-ray spectrum is dominated by a thermal soft X-ray component of temperature, $kT_{\rm eff} \sim 1$ keV, associated with a weak steep power-law component with photon index $\Gamma>$2.
In the low/hard state, on the other hand, spectra are mainly composed of Comptonized emission components in the hard X-ray band, soft excess components, and reflection components from the optically thick cold medium (for recent reviews, see McClintock \& Remillard 2003; Done et al. 2007).
While the high/soft state is reasonably well understood in terms of the standard disk model (see, e.g. Makishima et al. 1986), the nature of the accretion flow in the low/hard state is still a matter of debate.

Historically, two classes of models have been considered: the inner hot flow model and the disk corona model (Thorne \& Price 1975; Shapiro et al. 1976; Liang \& Price 1977; for a recent review of theoretical models, see Kato et al. 2008).
In the inner hot flow model, an optically thick disk is radially truncated at some radius, $r_{\rm tr}$, and the central region, $r<r_{\rm tr}$, is filled with a hot flow, which produces hard X-rays by means of the inverse Compton scattering of soft photons.
The properties of the hot flow are modeled by an advection-dominated accretion flow (ADAF, or more generally, radiatively inefficient accretion flow, RIAF), in which most of the energy released via viscous dissipation remains in the accreting gas rather than being radiated away (Ichimaru 1977;  Rees et al. 1982; Narayan \& Yi 1994).
The ADAF solution can exist only below a critical accretion rate, which depends on the viscosity parameter $\alpha$ (Ichimaru 1977; Narayan \& Yi 1995; Abramowicz et al. 1995).
Some authors calculated the radiative spectra from pure ADAFs or RIAFs by global Monte Carlo calculations (e.g. Kurpiewski \& Jaroszynski 1999; Kato et al. 2009).

The disk truncation is suggested by a number of observations.
The soft excess component in the low/hard state is generally attributed to soft thermal emission from optically thick, geometrically thin accretion disk (Ba\l uci\'{n}ska-Church et al. 1995; Ebisawa et al. 1996).
By spectral fitting of the soft thermal component to disk-blackbody (Mitsuda et al. 1984), the inner radius of the thin disk seems to be truncated at some outer radii, $r_{\rm tr} \gtsim 10$ Schwarzschild radii (Makishima et al. 2008; Gierli\'{n}ski et al. 2008).
Moreover the low/hard state spectrum shows a relatively small solid angle, $\Omega$, subtended by the reflector derived from the reflection component (Gierli\'{n}ski et al. 1997; Gilfanov et al. 1999).
This implies that the interface between the hot flow and the cool disk has rather small area.
One plausible explanation is that the inner part of the cool disk is truncated at some radius in the low/hard state.

The disk truncation is also suggested by the variability study.
The power density spectrum of the X-ray emission shows a typical break frequency (Miyamoto et al. 1992).
As the spectrum evolves harder, the frequency becomes smaller (Ibragimov et al. 2005; Axelsson et al. 2005).
This would imply that the inner radius of the disk is far from the radius of the marginally stable orbit in the low/hard state, given that the typical frequency is related to the motion of the inner edge of the disk.
 
In the disk corona model, by contrast, the optically thick disk remains untruncated, and hard X-rays are generated in a hot corona.
It is commonly assumed that a substantial fraction of the gravitational accreting energy is dissipated in the corona, although the physical processes by which the corona is heated are still unknown.
Magnetic fields amplified inside the disk could dissipate above the disk surface, where the density is relatively low, making an active hot corona (Galeev et al. 1979; Haardt \& Maraschi 1991; Di Matteo 1998; Merloni 2003).
The situation may be similar to the case of the solar corona.
Liu et al. (2003) solved the radial structure of magnetically heated corona and calculated the emergent spectra by summing up the radiation from the plane-parallel corona at each radius.
The corona could also be heated by the accretion of the coronal gas itself (Liu et al. 2002).
Kawanaka et al. (2009) studied the structure of the hot flow by using non-radiative MHD simulation data, identifying it as a corona.
They calculated self-consistent electron temperature and the emergent spectra by a global Monte Carlo calculation.

More recently, however, it is suggested by the Suzaku observations that the situation may be more complex.
Takahashi et al. (2008) and Makishima et al. (2008) found that the hard X-ray continua of GRO J1655-40 and Cyg X-1 are better fitted by two thermal Comptonization components, rather than by single component.
The existence of the two Comptonized components in the hard X-ray band has already been previously reported (Gierli\'{n}ski et al. 1997; Ibragimov et al. 2005).
These would imply a possibility that a hot gas generating hard X-ray photons has a complicated structure with multiple values of electron temperature and/or optical depth.
Another possibility of the origin of the two Comptonized components is a time-dependent effect (Poutanen \& Fabian 1999; Takahashi et al. 2008).
In this paper we consider the former case and neglect the properties of the time variability.

Though there are plenty of models for the hard X-ray spectra in the low/hard states, none of them can reproduce the two power-law components so far.
Especially the reproduction of the two power-law components is a theoretically challenging subject.
In this paper we investigate the structure and the emergent spectra of accretion flow for a variety of geometries of hot and cool accretion flows, instead of focusing on the detailed physical mechanisms making such accretion geometries, to examine which model can well explain the basic spectral properties of black holes during the low/hard state.
Here we consider steady models and neglect non-thermal electrons which would explain the MeV tail in the low/hard state spectra.
We will show that the combination of the inner hot flow model and disk corona model can naturally reproduce the two power-law components.
In section \ref{sec:model} we present our models used for the spectrum calculations and details of the numerical method.
Then we present the results of the emergent spectra in section \ref{sec:result}.
We discuss on the obtained results and applications for observation in section \ref{sec:discussions}.
Finally, we give conclusions of this paper in section \ref{sec:conclusions}.

\section{Our Models}
\label{sec:model}

\subsection{Model Requirements}
\label{sec:criteria}

In the present study, we aim at construction of accretion models which satisfy the basic spectral properties of the low/hard state.
For this purpose we set the following three criteria for successful models for the low/hard state spectra (Done et al. 2007; Makishima et al. 2008): \\
(i) The hard X-ray luminosity (at photon energy above 3keV), $L_{> 3{\rm keV}}$, should exceed the soft X-ray luminosity (below 3keV), $L_{< 3{\rm keV}}$.  \\
(ii) The hard X-ray luminosity should be in the range of $L_{> 3{\rm keV}}=0.003-0.2L_{\rm Edd} \simeq 0.4-30 \times 10^{37}{\rm \ erg \ s}^{-1}$ for black hole mass of $10 M_\odot$.  \\
(iii) The spectrum should exhibit two power-law components which are dominant above and below $\sim 10$keV, $\Gamma_{\rm s}  \sim 2 > \Gamma_{\rm h}$, where $\Gamma_{\rm s}$ (or $\Gamma_{\rm h}$) is the power-law index in the softer (harder) energy bands.

Other prominent properties of the low/hard state, e.g. MeV tales in the spectra and time-lags in Fourier space, are not considered in this paper.

\subsection{Various Geometries of the Accretion Flows}

\begin{figure}
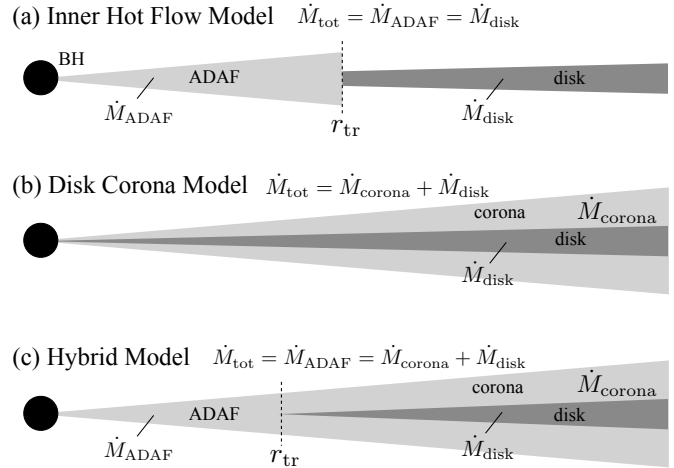

  \begin{center}
    \FigureFile(90mm,90mm){figure1.eps}
  \end{center}
  \caption{Schematic view of the three models of the accretion geometries.}
 \label{fig:geometry}
\end{figure}

We consider three representative models for the configuration of hot and cool flows (see figure \ref{fig:geometry}).
Generally speaking, hot and cool flows are expected to interact with each other by many different ways; e.g. energy exchange via radiation, conduction, and magnetic fields, mass exchange via evaporation or condensation, and  so on.

Here, we focus on the radiative interaction of the flows and do not explicitly consider thermal conduction nor solve the dynamics of magnetic fields.
Though there are some attempts for calculating the mass evaporation/condensation process between the hot flow and the cool thin disk (e.g., R\'{o}\.{z}a\'{n}ska \& Czerny 2000; Spruit \& Deufel 2002; Liu et al. 2007; Qiao \& Liu 2009), we do not follow them since there exist uncertainties in the treatments, but instead incorporate the effects of matter exchange by changing the accretion rate within the cool and hot flows.
The exchanges of the energy and angular momentum between the two phases 
(other than the energy exchange via radiation) are not considered in the present study. 
The effects of wind mass loss from the hot flow will be discussed in section \ref{sec:discussions}.

The basic assumptions for each of the three models are as follows:
\\
Model (a): Inner Hot Flow Model

In this model we postulate that a standard thin disk surrounds an inner hot ADAF.
We assume that the material in the outer thin disk suddenly evaporates to the ADAF 
at the transition radius, $r=r_{\rm tr}$,
so that the transition zone between the outer thin disk and the inner ADAF is sufficiently narrow.
The accretion rate of the inner ADAF, $\dot{M}_{\rm ADAF}$, is assumed to be the same as that of the outer thin disk, $\dot{M}_{\rm disk}$, and these values are assumed to be radially constant, 
$\dot{M}_{\rm tot}=\dot{M}_{\rm ADAF}=\dot{M}_{\rm disk}={\rm constant}$, where $\dot{M}_{\rm tot}$ is the total accretion rate of the hot flow and cool flow.
Some fraction of the soft photons emitted from the thin disk is captured by the ADAF and cools the ADAF, especially near the transition radius,
but only a small fraction of the hard photons generated by Compton up-scattering in the ADAF is caught by the outer thin disk because of the small cross section of the thin disk.
The reprocessed radiation is thus expected to be weak.
\\
Model (b): Disk Corona Model

In this model we consider two-temperature, accreting coronae above and below a thin cool disk.
The corona is powered by the release of the gravitational energy by accretion itself.
The thin disk extends down to the radius of the innermost stable circular orbit (ISCO) of the non-spinning black hole space-time, $r = 3 r_{\rm S}$.
We assume that the accretion rate of the corona, $\dot{M}_{\rm corona}$, and that of the underneath thin disk, $\dot{M}_{\rm disk}$, are both constant in space.
Since the hot corona covers the whole thin disk, we expect that most of soft photons emitted from the thin disk experience scattering in the coronae and that a large fraction of the hard photons generated in the coronae are reprocessed by the thin disk.
This makes a big contrast with the case of the inner hot flow model (a).
\\
Model (c): Hybrid Model

We finally consider a hybrid model of Models (a) and (b).
The outer part has a disk-corona structure, which is, however, truncated at some radius, 
$r=r_{\rm tr}$, and the flow becomes pure ADAF inside this radius.
The mass accretion rate of the inner ADAF is a sum of that of the outer corona and the thin disk, 
i.e., $\dot{M}_{\rm ADAF}=\dot{M}_{\rm tot} \equiv \dot{M}_{\rm corona}+\dot{M}_{\rm disk}$. 
We assume that these values are constant in space unless specified otherwise.
Note that similar configurations were considered in some previous studies, though the structure of the hot flow was assumed to be uniform (e.g. Poutanen et al. 1997).

\subsection{Basic assumptions and simplifications}

In order to calculate the emergent spectra, we need to solve the radial structure of the ADAF, the corona, and the thin disk.
Note that what matters in the thin disk part is only the radial profile of the soft photon emissivity
(or more precisely, the energy dissipation rate as a function of the radius) and it can be easily derived from the standard disk theory once an accretion rate is specified.
As for the structure of the ADAF and corona we adopt the technique used for the self-similar solution of the advection-dominated flow (Narayan \& Yi 1995).  
Note, however, that the ADAF and the corona show somewhat different temperature profiles since the amount of soft radiation available for Compton cooling differs.

In a realistic situation, dynamics of the magnetic fields and the way of interactions with the thin disk (via, e.g., thermal conduction and/or convective motions) could be markedly different in the corona and the ADAF.
These issues are beyond the scope of the present study and are left as future issues.

Throughout the present study we adopt the Newtonian dynamics and assume an axisymmetric and steady flow structure in the Newtonian gravitational potential, $\psi = -GM/r$, where $M$ is the black hole mass.
We neglect the radiative force on the ADAF/corona, since the hot flow is optically thin and the typical luminosity is sub-Eddington.
We further assume that the accreting gas is composed of two-temperature plasmas of fully ionized hydrogen. 
Electron-positron pairs are not considered.

Here the blackbody emission from the thin disk is assumed to be the source of soft photons and we neglect seed photons generated by synchrotron emission.
Note that the synchrotron emission would be weak in a pure thermal plasma (Wardzi\'{n}ski \& Zdziarski 2000).
However the synchrotron emission would provide enough soft photons for the low/hard spectrum, if MeV(non-thermal) tail exists in the electron distribution, which is required in order to explain the observed spectrum in MeV band. (Wardzi\'{n}ski \& Zdziarski 2001; Malzac \& Belmont 2009;  Poutanen \& Vurm 2009). 

\subsection{Basic Equations for Radial Structure of Hot Accretion Flow}

We employ the vertically integrated equations.
Assuming the hydrostatic balance in the vertical direction, we obtain the scale hight of the ADAF/corona, 
\begin{eqnarray}
H = r \frac{c_{\rm s}}{v_{\rm K}}, 
\end{eqnarray}
where $c_{\rm s} \equiv \Pi/\Sigma$ is the sound speed, and $\Sigma$ and $\Pi$ is vertically integrated gas density and pressure, respectively.
We assume vertically uniform structure of the ADAF/corona within the half thickness, $H$. 
Hence the surface density is given by $\Sigma = 2 H m_{\rm p} n$, where $n$ is the number density of protons, and the mass accretion rate of the ADAF(or corona) is $\dot{M}_{\rm ADAF} (\dot{M}_{\rm corona}) \equiv - 2 \pi r \Sigma v_r$.

We obtain azimuthal velocity from the radial momentum equation,
\begin{eqnarray}
v_\varphi^2 = v_\mathrm{K}^2 + \frac{1}{\Sigma} \frac{d \Pi}{d \ln r}, \label{eq:radial}
\end{eqnarray}
where $v_{\rm K}$ is Keplerian velocity and the last term is radial pressure gradient.
In this paper we neglect radial inertial term for simplicity.
It should be noted that this term contributes at most $\sim 30 \%$ of the $v_{\rm K}^2$ in equation (\ref{eq:radial}) even for $\alpha = 1$, which we adopt for reproducing the observational results.

The angular momentum equation is 
\begin{eqnarray}
\Sigma v_r \frac{d}{dr} (r v_\varphi) = \frac{1}{r} \frac{d}{dr} (r^2 T_{r \varphi}),
\end{eqnarray}
where we assume that the $r$-$\varphi$ component of  the viscous stress tensor, $T_{r \varphi}$, is dominant and other components are negligible.
We make the Ansatz that the stress tensor scales with the local gas pressure with a constant of proportionality, $T_{r \varphi}=-\alpha \Pi$.
In order to solve the dynamical properties of the flow, we assume that the ADAF and the corona have self-similar structures, $\Sigma \propto r^{-1/2}$ and $\Pi \propto r^{-3/2}$.

We assume that the electrons and protons are thermodynamically coupled only through Coulomb coupling, neglecting poorly understood direct heating processes of electrons via magnetic fields.
The energy balance between protons and electrons are, respectively,
\begin{eqnarray}
\frac{1}{\gamma-1} v_r \frac{d \Pi}{dr} - \frac{\gamma}{\gamma-1} v_r \frac{\Pi}{\Sigma} \frac{d \Sigma}{dr}+Q_{\rm Coulomb} &=& r T_{r \varphi} \frac{d}{dr} \left( \frac{v_\varphi}{r} \right), \label{eq:proton} \\
Q_\mathrm{Comp} + Q_\mathrm{brems} &=& Q_\mathrm{Coulomb}, \label{eq:electron}
\end{eqnarray}
where the first two terms on left-hand-side of equation (\ref{eq:proton}) represent advective cooling and the term on right-hand-side is viscous heating.
Here an additional heating at the transition radius is not considered (Manmoto \& Kato 2000).
We also assume that an advective term of electrons is negligible (see Nakamura et al. 1997).
This means that the electron cooling timescale is much shorter than the accretion timescale, $r/v_r$.
We find that this assumption is valid unless the truncation radius is very large $(\gtrsim 100 r_{\rm S})$.

The vertically integrated pressure of the fully ionized gas is 
\begin{eqnarray}
\Pi = \frac{\Sigma k T_{\rm p}}{m_{\rm p}},
\end{eqnarray}
where $T_{\rm p}$ is the temperature of protons and we neglect the pressure of electrons since their temperature is sufficiently low, $T_{\rm e} \ll T_{\rm p}$.

In our Newtonian model, the gas is assumed to be immediately swallowed by the black hole, once it reaches the inner boundary of the flow at $3 r_{\rm S}$, thereby emitting no radiation from inside this radius.
In order to save the calculation time we restrict the computational region, $r \le r_{\rm out}$, where $r_{\rm out}$ is the outer boundary radius and $r_{\rm out} = 100 r_{\rm S}$ unless stated otherwise.
We employ logarithmically spaced 100 grid points to describe the properties of the ADAF, the corona, and the thin disk.  
The basic equations can be integrated when cooling rates are calculated (see the next subsections).

\subsection{Coulomb Coupling and Bremsstrahlung Cooling}

The energy transport rate from protons to electrons by Coulomb collision is (Stepney \& Guilbert 1983)
\begin{eqnarray}
Q_\mathrm{Coulomb} = 2 H \times \frac{3}{2} \frac{m_{\rm e}}{m_{\rm p}} n^2 \sigma_\mathrm{T} c \frac{kT_{\rm p}-kT_{\rm e}}{K_2(1/\Theta_{\rm e}) K_2(1/\Theta_{\rm p})} \ln \Lambda \nonumber \\ 
 \times \left[ \frac{2(\Theta_{\rm e} + \Theta_{\rm p})^2 + 1}{\Theta_{\rm e} + \Theta_{\rm p}}  K_1 \left( \frac{\Theta_{\rm e} + \Theta_{\rm p}}{\Theta_{\rm e} \Theta_{\rm p}}  \right) + 2 K_0 \left( \frac{\Theta_{\rm e} + \Theta_{\rm p}}{\Theta_{\rm e} \Theta_{\rm p}}  \right) \right],
\end{eqnarray}
where $\Theta_{\rm p}$ and $\Theta_{\rm e}$ are the temperatures of the protons and electrons normalized by their rest mass energy, $\Theta_{\rm p} \equiv k T_{\rm p}/m_{\rm p} c^2$ and $\Theta_{\rm e} \equiv k T_{\rm e}/m_{\rm e} c^2$.
Here $\sigma_{\rm T}$ is the cross section of the Thomson scattering, $\ln \Lambda$ is the Coulomb logarithm, $\ln \Lambda \sim 20$, and $K_i$ is modified Bessel function of the second kind of order $i$.

The bremsstrahlung cooling rate per unit area is
\begin{eqnarray}
Q_\mathrm{brems} = 2 H \times \alpha_\mathrm{f} \sigma_\mathrm{T} m_{\rm e} c^3 n^2 F(\Theta_{\rm e}),
\end{eqnarray}
where $\alpha_\mathrm{f}$ is the fine-structure constant.
Here the dimensionless factor is (Svensson 1984; Kato et al. 2008)
\begin{eqnarray}
F(\Theta_{\rm e}) =  \cases{
2 \left( \frac{2}{\pi} \right)^{3/2} \Theta_{\rm e}^{1/2} \left( 1 + 1.78 \Theta_{\rm e}^{1.34} \right) \cr
+ \frac{5}{6 \pi^{3/2}} (44-3\pi^2) \Theta_{\rm e}^{3/2} \left( 1+1.1\Theta_{\rm e} + \Theta_{\rm e}^2 - 1.25 \Theta_{\rm e}^{2.5} \right), \cr
 {\rm \ \ \ \ \ \ \ \ \ \ \ \ \ \ \ \ \ \ \ \ \ \ \ \ \ \ \ \ \ \ \ \ \ \ \ \ \ \ \ \ \ \ for\ }  \Theta_{\rm e} \le 1;\cr
\frac{9}{2 \pi} \Theta_{\rm e} \left[ \ln \left( 2 e^{-\gamma_\mathrm{E}} \Theta_{\rm e} + 0.42 \right) +\frac{3}{2} \right] \cr
+ \frac{9}{\pi} \Theta_{\rm e} \left[ \ln \left( 2 e^{-\gamma_\mathrm{E}} \Theta_{\rm e} \right) +\frac{5}{4} \right],
 {\rm \ \ \ \ \ \ \ \ for \ } \Theta_{\rm e} > 1,\cr}
\end{eqnarray} 
where $\gamma_\mathrm{E} \simeq 0.5772$ is Euler's number.
In our model we do not consider a supply of the soft photons from the bremsstrahlung emission in the Comptonization process, since the emission rate is sufficiently smaller than that of soft photons from the thin disk.

\subsection{Reprocessing of Comptonized Photons by the Thin Disk}
\label{sec:reprocess}

The dissipation rate of energy in the thin disk is 
\begin{eqnarray}
Q_{\rm disk} = \frac{3}{4 \pi} \frac{GM \dot{M}_{\rm disk}}{r^3} \left( 1-\sqrt{\frac{r_{\rm in}}{r}} \right),
\end{eqnarray}
where $r_{\rm in}=3 r_{\rm S}$ is the radius of the inner edge of the disk, where we adopt a torque-free condition.

We take into account the radiative coupling between the corona and the thin disk by including reprocessing of radiation by the thin disk, as described by Haardt \& Maraschi (1991).
Some part of photons up-scattered in the ADAF/corona returns to the thin disk and heats up the surface of the disk.
Hence the soft photon emission rate per unit area is the sum of the energy generated in the disk, $Q_{\rm disk}$, and the reprocessed radiation, $Q_{\rm rep}$,
\begin{eqnarray}
Q_{\rm soft} = Q_{\rm disk} + Q_{\rm rep},
\end{eqnarray}
where we set the albedo of the thin disk $a=0$, i.e., all the photons impinging on the disk are assumed to be absorbed.
This soft photon emission cools down the electrons in the ADAF/corona by inverse Compton scattering.

\subsection{Monte Carlo Calculation}
\label{sec:Monte}

We calculate the radiative spectra emerging from the accretion flow structures of the three geometries described above.
We consider a computational region of a cylinder with a radius of $r_{\rm out}$ and a height of $2 r_{\rm out}$.
We neglect the height of the thin disk; we assumed that photons should be absorbed by the thin disk, if the path of the photon crosses the equatorial plane, $z=0$, at $r>r_{\rm tr}$.
We also neglect the absorption of the photon by free-free process.
The bending of photon trajectory near the black hole nor absorption of photons by black hole event horizon are not included in this calculation. 
Inside the inner boundary of the hot flow, $r<3 r_{\rm S}$, we neglect any scattering. (In the computational procedure we set sufficiently large mean free path, $\lambda \sim 10^{10}  r_{\rm S}$, and sufficiently low electron temperature, $T_{\rm e} \sim 10$ eV.)

We calculated the photon spectra scattered by hot electrons by using Monte Carlo calculation, following Pozdnyakov, Sobol, \& Sunyaev (1977) and Liu, Mineshige, \& Ohsuga (2003). 
We first choose the position of initial soft photon, $r=r_0$, from the radial distribution of the soft photon emissivity from the thin disk, $Q_{\rm soft}(r)$.
We set the weight $w_0=1$ for an initial photon with energy $E_0$, which has Planck distribution with temperature at $r_0$. 
We calculate the probability $P_0= \exp(-\tau)$ for photon passing through the hot gas, where $\tau$ is optical depth evaluated from the photon position to the boundary of computational region or the thin disk. 
Then $w_0 P_0$ is the transmitted portion of photon and the remaining $w_1 = w_0 (1-P_0)$ is the portion of photon scattered at least once.
Let $w_n = w_{n-1} (1-P_{n-1})$ be the portion of photon experiencing the $n$-th scattering. 
We continue the calculation until the weight becomes sufficiently small, $w_n < 10^{-7}$. 
Repeating the same procedures for sufficiently large number of photons, typically $10^{8}$, we calculate emergent spectra by collecting photons escaping from the computational region and reprocessed radiation by collecting photons absorbed by the thin disk.
We also calculate the cooling rate of the electrons at each radial bin by evaluating the change of the photon energy at every scattering.
In this paper we do not calculate the reflection features from the thin disk.

By this Monte Carlo calculation, we obtain the quantities of radiative processes, $Q_{\rm Comp}$ and $Q_{\rm rep}$.
In order to obtain self-consistent structures of accretion flows, we use these $Q_{\rm Comp}$ and $Q_{\rm rep}$ values to re-calculate the flow structure and iterate (see below).

\subsection{Iterative Method}
\label{sec:iteration}

We first solve the structure of the ADAF and the corona by using tentative values of the Compton cooling rate, $Q_{{\rm Comp}, 0}$, and the emissivity of reprocessed photons, $Q_{{\rm rep}, 0}$, where the subscripts denote the number of iterations.
Then we solve the radiative transfer by the Monte Carlo calculation, obtaining the exact values, $Q_{{\rm Comp}, 1}$ and $Q_{{\rm rep}, 1}$, for this structure.
We solve the structure of the flow again by using the newly obtained cooling rate, $Q_{{\rm Comp}, 1}$, and calculate the radiative spectra by using $Q_{{\rm rep}, 1}$.
Repeating the same procedure, we finally obtain the self-consistent structure and radiative spectra.

\section{Results of the X-ray Spectra}
\label{sec:result}

There are four main model parameters: the viscosity parameter $\alpha$, the total mass accretion rate $\dot{m}_{\rm tot}$, $\dot{m}_{\rm corona}/\dot{m}_{\rm disk}$, and the truncation radius $r_{\rm tr}$, where $\dot{m}_{\rm tot} \equiv \dot{M}_{\rm tot} c^2 /L_{\rm Edd}$, $\dot{m}_{\rm corona}\equiv \dot{M}_{\rm corona} c^2 /L_{\rm Edd}$, and $\dot{m}_{\rm disk} \equiv \dot{M}_{\rm disk} c^2 /L_{\rm Edd}$.
For clarity we show the typical values of parameters used in this paper: $\alpha=$0.3 and 1, $\dot{m}_{\rm tot}=$0.1, 0.3, 0.8, and 1, $\dot{m}_{\rm corona}/\dot{m}_{\rm disk}=1-1000$, and $r_{\rm tr}/r_{\rm S}=3-300$.
Note that relatively large values of $\alpha$ is necessary to obtain the advection-dominated solutions at large accretion rates (i.e., hard X-ray luminosities).
The black hole mass, $m \equiv M/M_\odot$, should be adjusted to each object.
Hereafter we study the cases of the stellar mass black holes, $m=10$.

\subsection{(a) Inner Hot Flow Model}
\label{sec:inner}

In this model the mass accretion rate of the ADAF, $\dot{m}_{\rm ADAF} \equiv \dot{M}_{\rm ADAF} c^2/L_{\rm Edd}$, is assumed to be the same as that of the thin disk.
Note that $\dot{m}_{\rm ADAF}$ cannot be larger than the critical value, which depends on the viscosity parameter $\alpha$.

\begin{figure*}
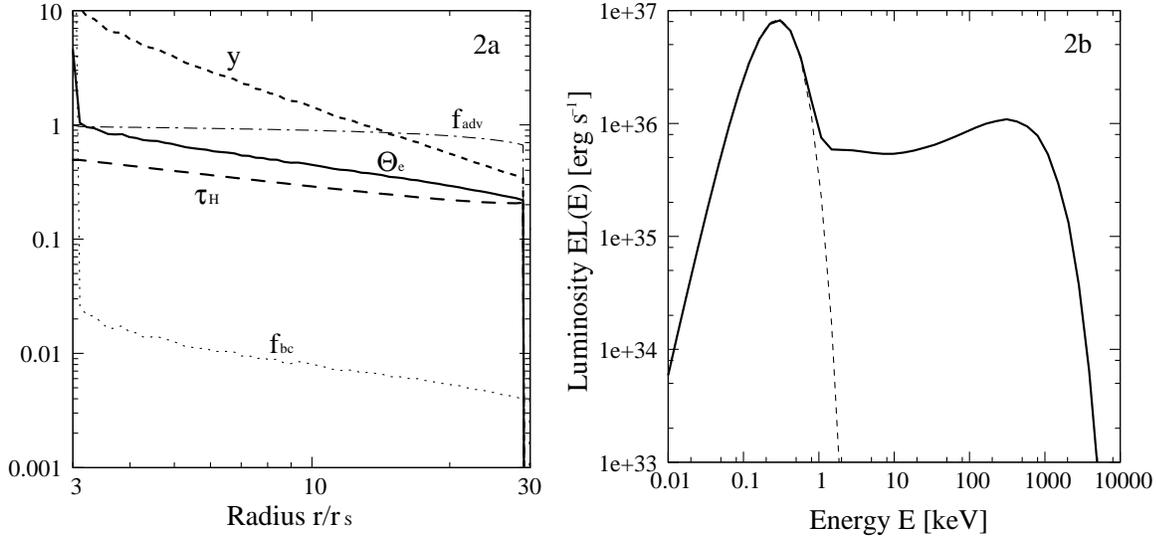

  \begin{center}
    \FigureFile(73mm,73mm){figure2a.eps}
    \FigureFile(80mm,80mm){figure2b.eps}
  \end{center}
  \caption{The structure of the ADAF (\ref{fig:inner1}a: left panel) and the emergent spectra (\ref{fig:inner1}b; right panel) for $\alpha=1$, $\dot{m}_{\rm ADAF}= 1$, and $r_{\rm tr}=30 r_{\rm S}$. Left panel (\ref{fig:inner1}a): In the ADAF region, $3 r_{\rm S} < r < r_{\rm tr}$, we show the dimensionless values, electron temperature, $\Theta_{\rm e}$ (solid line), vertical optical depth, $\tau_{\rm H}$(long-dashed line), Compton $y$-parameter (short-dashed line), advection fraction, $f_{\rm adv}$(dash-dotted line), and the ratio of the cooling rate of bremsstrahlung emission to the Compton cooling rate, $f_{\rm bc}$(dotted line). 
Right panel (\ref{fig:inner1}b): The spectrum of the photons escaping from the computational region and soft photons from the thin disk is shown by solid and dashed line, respectively. 
}
 \label{fig:inner1}
\end{figure*}

We calculate the radial structure of the ADAF and show the results for $\alpha=1$, $\dot{m}_{\rm ADAF}= 1$, and $r_{\rm tr}=30 r_{\rm S}$ in figure \ref{fig:inner1}a.
Here the vertical optical depth is $\tau_{\rm H} \equiv n \sigma_{\rm T} H$, and the Compton $y$-parameter is $y \equiv (4 \Theta_{\rm e} + 16 \Theta_{\rm e}^2) \tau_{H}$.
The advection fraction, $f_{\rm adv} \equiv Q_{\rm adv}/Q_{\rm visc}$, is approximately unity, validating the assumption that the hot flow is advection-dominated.
This means that the Coulomb coupling timescale is much longer than the accretion timescale.
Hence the scale height and the optical depth follows the advection-dominated solutions, $H \sim r$ and $\tau_{\rm H} \propto \alpha^{-1} \dot{m}_{\rm ADAF} r^{-1/2}$ as long as $f_{\rm adv} \sim 1$ (Narayan \& Yi 1995).
The ratio of the bremsstrahlung emission rate to the Compton cooling rate, $f_{\rm bc} \equiv Q_{\rm brems}/Q_{\rm Comp}$, is so small that we can safely neglect the contribution of the bremsstrahlung emission to the spectrum.

We display the emergent spectra from this structure in figure \ref{fig:inner1}b.
Here we do not consider the reflection feature in the spectrum, which is not expected to be so strong.
Since the Compton cooling is most efficient at $r=r_{\rm tr}$, where the emission of the soft photons from the thin disk is largest, the electron temperature decreases with radius.
Moreover the optical depth increases with decreasing radius, so does the $y$-parameter.
Since the soft photons come only from the thin disk part located at $r>r_{\rm tr}$, a small fraction of the soft photons is intercepted by the inner ADAF, resulting in a very weak Compton component in the radiative spectra (about $30 \%$ of the thermal component; see Table 1), which  does not  agree with the observations.
Note that we only consider the soft photons from the thin disk inside the boundary $r \le r_{\rm out}=100r_{\rm S}$.
Thus we slightly underestimate the soft X-ray luminosity since we neglect the soft photons from the thin disk at $r>r_{\rm out}$ (see below).

We calculate variety of the emerging spectra for other parameters (Table 1).
We find that the hard X-ray component is always much weaker than the soft X-ray component.
As $r_{\rm tr}$ decreases, the ratio of the hard X-ray luminosity to the soft X-ray luminosity, $L_{\rm > 3keV}/L_{\rm < 3keV}$, decreases, though the most of the energy is dissipated in the inner part of the ADAF.
This is because an increase in the number of soft photons results in decreasing electron temperature  due to more efficient Compton cooling of the ADAF region.

\begin{table*}
\begin{center}
\caption{Parameters and the resultant values for the inner hot flow model}
 \begin{tabular}{ccccccc}
  \hline
  \multicolumn{1}{c}{$\alpha$}  & $\dot{m}_{\rm ADAF}$ & $r_{\rm tr}/r_{\rm S}$ &  $L_{<3 {\rm keV}}$ [${\rm erg \ s^{-1}}$]\footnotemark[$*$] & $L_{>3 {\rm keV}}$ [${\rm erg \ s^{-1}}$]\footnotemark[$*$] & ratio\footnotemark[$*$] & $\Gamma$\footnotemark[$\dagger$]  \\
  \hline
  1 & 1 & 10 & $4.2 \times 10^{37}$ & $7.2 \times 10^{36}$ & 0.17 & 2.05 \\
  1 & 1 & 30 & $1.4 \times 10^{37}$ & $4.6 \times 10^{36}$ & 0.33 & 1.88 \\
  1 & 1 & 30\footnotemark[$\ddagger$] & $2.0 \times 10^{37}$ & $4.8 \times 10^{36}$ & 0.23 & 1.88 \\
  1 & 1 & 100\footnotemark[$\ddagger$] & $6.2 \times 10^{36}$ & $2.0 \times 10^{36}$ & 0.32 & 1.88 \\
  1 & 1 & 300\footnotemark[$\ddagger$] & $1.7 \times 10^{36}$ & $6.1 \times 10^{35}$ & 0.35 & 1.70 \\
  1 & 0.3 & 30 &  $3.9 \times 10^{36}$ & $2.3 \times 10^{35}$ & 0.06 & 2.11 \\
  0.3 & 0.1 & 30 &  $1.3 \times 10^{36}$ & $2.3 \times 10^{35}$ & 0.17 & 1.98 \\
  \hline
  \multicolumn{4}{@{}l@{}}{\hbox to 0pt{\parbox{120mm}{\footnotesize
         \par\noindent
         \footnotemark[$*$] The integrated luminosity below and over 3 keV, and the ratio $L_{>3 {\rm keV}}/L_{<3 {\rm keV}}$.
         \par\noindent
         \footnotemark[$\dagger$] Photon index evaluated by two points of 3 and 100 keV in the spectra.
         \par\noindent
         \footnotemark[$\ddagger$] $r_{\rm out}=1000r_{\rm S}$.
   }\hss}}
  \end{tabular}
\end{center}
\label{tab:modela}
\end{table*}

We also calculate the radiative spectra for much larger values of the truncation radius than that estimated by the observational results, $r_{\rm tr} \sim 10 r_{\rm S}$ (see e.g. Makishima et al. 2008).
In such cases we should set large outer boundary $r_{\rm out}=1000r_{\rm S}$ in order not to underestimate the soft photon luminosity from the outer thin disk.
We see that the radio of the hard X-ray luminosity to the soft X-ray luminosity is still small ($\lesssim 0.3$) for large values of $r_{\rm tr}$ (Table 1).
To make the situation worse, the luminosity of hard X-ray, as well as soft X-ray, decreases as the truncation radius increases.
This is because, the larger $r_{\rm tr}$ is, the smaller becomes the number of the soft photons entering the inner region of the ADAF and the less efficient becomes Compton cooling.
Then the electron temperature becomes higher and the energy transfer rate from protons to electrons by Coulomb collisions (which is approximately proportional to $T_{\rm e}^{-3/2}$) becomes smaller (Coulomb coupling timescale becomes longer).
Consequently the hard X-ray luminosity decreases (advection fraction increases) as $r_{\rm tr}$ increases.
The hard X-ray luminosity in the case of $r_{\rm tr}=300 r_{\rm S}$ is one order of magnitude smaller than the criterion for the low/hard state.
Note that for large $r_{\rm tr}$($\gtrsim 300$) the bremsstrahlung emission from the inner region of the ADAF dominates over the Comptonized radiation.
To summarize, we cannot reproduce the X-ray spectra in the low/hard state by the inner hot flow model for $(\alpha, \dot{m}_{\rm ADAF})=(1,1)$.

For a given $\alpha$, the hard X-ray component becomes harder and stronger as the accretion rate, $\dot{m}_{\rm ADAF}$ increases.
However, the mass accretion rate cannot be larger than the critical value, $\dot{m}_{\rm crit} \sim 1$(or 0.1) for $\alpha=1 (0.3)$ in the inner ADAF.
We can see that even if the accretion rate is its maximum value for a given $\alpha$, the hard X-ray component is still weaker than the soft X-ray component (see the case of $\alpha=0.3$ in Table 1).

In summary the inner hot flow model always produces too weak hard X-rays to explain the observed spectra in the low/hard state.

\subsection{(b) Disk Corona Model}
\label{sec:corona}

\begin{figure*}
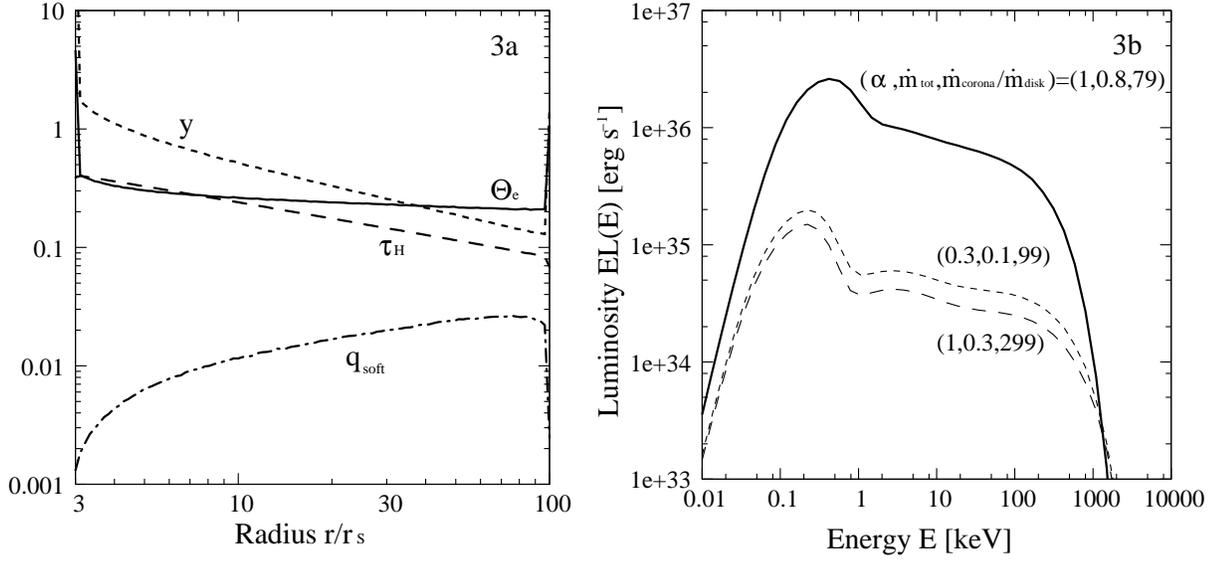

  \begin{center}
    \FigureFile(77mm,77mm){figure3a.eps}
    \FigureFile(83mm,83mm){figure3b.eps}
  \end{center}
  \caption{The structure of the corona (\ref{fig:corona}a; left panel) and the emergent spectra (\ref{fig:corona}b; right panel). Left panel (\ref{fig:corona}a): We show $\Theta_{\rm e}$(solid line), $\tau_{\rm H}$(long-dashed line), and $y$(short-dashed line), as well as the soft photon emissivity, $q_{\rm soft}$(dash-dotted line) for $(\alpha, \dot{m}_{\rm tot}, \dot{m}_{\rm corona}/\dot{m}_{\rm disk})$=(1, 0.8, 79). Right panel (\ref{fig:corona}b): Emergent spectrum for $(\alpha, \dot{m}_{\rm tot}, \dot{m}_{\rm corona}/\dot{m}_{\rm disk})$=(1, 0.8, 79) (solid line), (1, 0.3, 299)(long-dashed line), and (0.3, 0.1, 99)(short-dashed line).}
 \label{fig:corona}
\end{figure*}

In this model we assign three parameters: the viscosity parameter $\alpha$, total accretion rate $\dot{m}_{\rm tot} = \dot{m}_{\rm corona}+\dot{m}_{\rm disk}$, and the ratio of the accretion rate of the corona to that of the thin disk $\dot{m}_{\rm corona}/\dot{m}_{\rm disk}$.
We show the structure of the coronal flow for $(\alpha, \dot{m}_{\rm tot}, \dot{m}_{\rm corona}/\dot{m}_{\rm disk})$=(1, 0.8, 79) in figure \ref{fig:corona}a.
The maximum accretion rate for $\alpha=1$ is found to be $\dot{m}_{\rm tot} \sim 0.8$ in the disk corona model. 
The electron temperature is almost constant in space, in contrast to the case of the inner hot flow model.
This is because significant Compton cooling in the innermost region by the underlying thin disk where emissivity rapidly increases inward.
Nevertheless the inner region has the largest $y$-parameter, because of a large optical depth.
Note that the electron temperature increases at the inner edge, since the emission rate of the soft photons, $q_{\rm soft} \equiv r^2 Q_{\rm soft}/\dot{M}_{\rm tot} v_{\rm K}^2$, vanishes at the torque free radius.

The solid line in figure \ref{fig:corona}b shows the result of the spectrum from the coronal structure described above.
The emergent spectrum has a single power-law component in the hard X-ray band, though the $y$ parameter varies in the radial direction.
We find that the hard X-ray luminosity is only $\sim 40 \%$ of the soft X-ray luminosity, and the photon index is $\Gamma \sim 2.2$.
Hence the spectrum does not satisfy the criteria for the low/hard state, even when a large ratio of $\dot{m}_{\rm corona}/\dot{m}_{\rm disk}$ is adopted.

For small mass accretion rate, $\dot{m}_{\rm tot}=0.3$, the resulting spectrum has two humps in the hard X-ray band, rather than a single power-law component (see figure \ref{fig:corona}b).
This is because the electron temperature is so high, $\Theta_{\rm e} \sim 1$, that the photons gain significant energies only by a single scattering.
We can see that the hump around 10 keV is made by one scattering of the soft photons, and another high energy hump is made by the photons scattered at least twice.
We also show the emergent spectra for $(\alpha, \dot{m}_{\rm tot})=(0.3, 0.1)$ in figure \ref{fig:corona}b.
We find that its spectral features are very similar to those for $(\alpha, \dot{m}_{\rm tot})=(1, 0.3)$.
Note that we cannot find the advection-dominated solution of the corona with larger accretion rate, $\dot{m}_{\rm tot}>0.1$, for $\alpha=0.3$.

\begin{figure*}
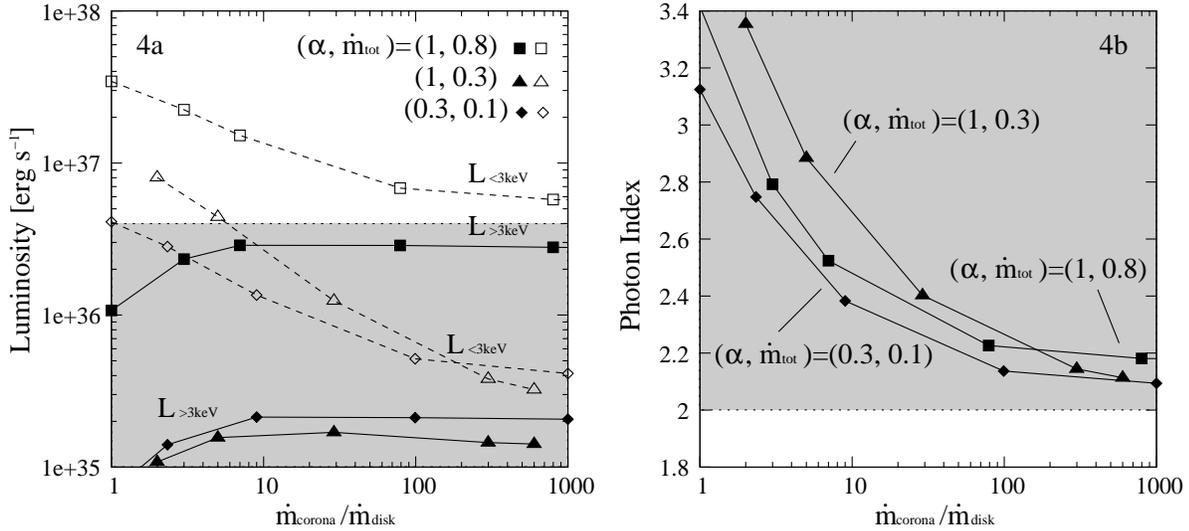

  \begin{center}
    \FigureFile(80mm,80mm){figure4a.eps}
    \FigureFile(77mm,77mm){figure4b.eps}
  \end{center}
  \caption{The spectral features of the disk corona model as functions of the ratio of the accretion rate of the corona to that of the thin disk, $\dot{m}_{\rm corona}/\dot{m}_{\rm disk}$, for given $\alpha$ and $\dot{m}_{\rm tot}$.
Squares, triangles, and diamonds represent the results for $(\alpha, \dot{m}_{\rm tot}) = (1, 0.8)$, $(\alpha, \dot{m}_{\rm tot}) = (1, 0.3)$, and $(\alpha, \dot{m}_{\rm tot}) = (0.3, 0.1)$ respectively.
Left panel (\ref{fig:corona2p}a): The luminosity of the emergent spectra below and over 3keV, $L_{\rm < 3 keV}$ (open symbols) and $L_{\rm > 3 keV}$ (filled symbols).
The spectrum satisfies the criterion for $L_{\rm > 3 keV}$ when the filled symbol is above the shaded region.
Right panel (\ref{fig:corona2p}b): The photon index evaluated by the two points, 3keV and 100keV.
The photon index should be below the shaded region in order to satisfy the criterion.}
 \label{fig:corona2p}
\end{figure*}

We show in figure \ref{fig:corona2p} (squares) the resultant luminosities and photon indices for various ratios of the accretion rate $\dot{m}_{\rm corona}/\dot{m}_{\rm disk}$ but keeping $\dot{m}_{\rm tot}=1$.
We see that the spectrum has harder (smaller $\Gamma$) and stronger power-law component for larger ratios of $\dot{m}_{\rm corona}/\dot{m}_{\rm disk}$.
However the hard X-ray luminosity is less than $50 \%$ of the soft X-ray luminosity and the photon index is larger than 2, even when $\dot{m}_{\rm corona}/\dot{m}_{\rm disk}=1000$.

The situation is not improved but rather getting worse for the cases with a smaller $\dot{m}_{\rm tot}=0.3$ or with a smaller $\alpha=0.3$ (see triangles and diamonds in figure \ref{fig:corona2p}).
In the both cases, the hard X-ray luminosity is much smaller than the observed values, $L_{\rm >3 keV} \lesssim 2 \times 10^{35}$ erg s$^{-1}$.

For the coronal geometry we expect that the fraction of the escaping photons from the corona should be comparable to that of the reprocessed photons. 
Hence, the thermal component should exceed the power-law component.
Similar results were obtained previously by an one-zone disk corona model (e.g. Haardt \& Maraschi 1993; Stern et al. 1995; Poutanen et al. 1997; Dove et al. 1997; Cao 2009).
We confirm these results by the radially structured corona models.
We conclude that the hard X-ray component can never exceed the soft X-ray component even for large $\dot{m}_{\rm tot}$ and large $\dot{m}_{\rm corona}/\dot{m}_{\rm disk}$ and the photon index of the power-law component is always lager than two.
The situations will be improved, if the thin disk is truncated (see below).

\subsection{(c) Hybrid Model}
\label{sec:hybrid}

\begin{figure*}
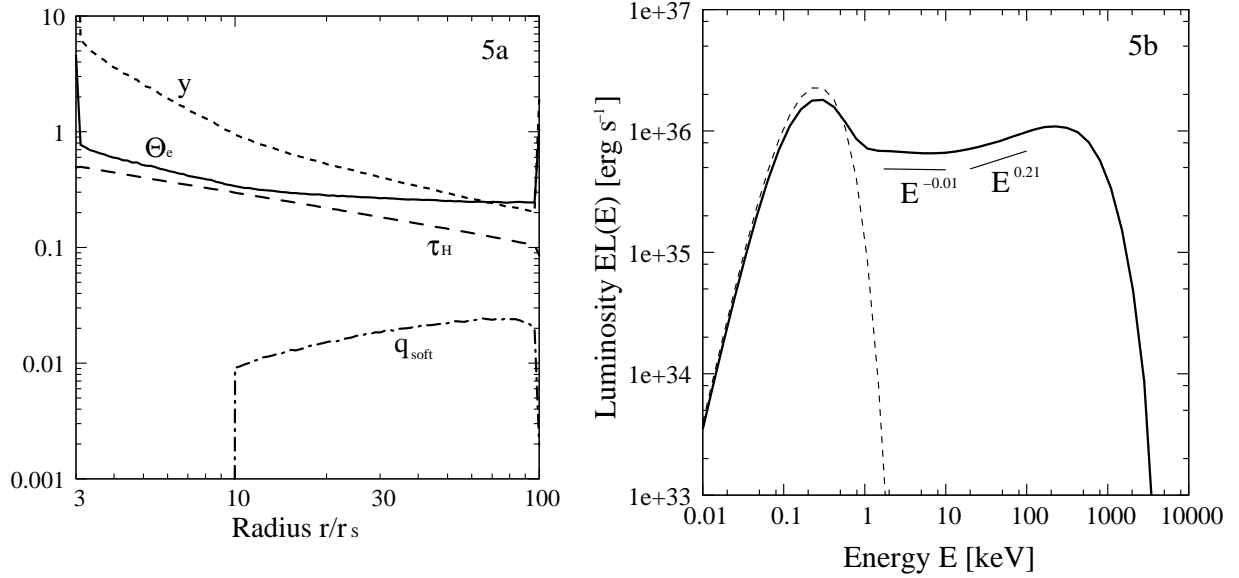

  \begin{center}
    \FigureFile(77mm,77mm){figure5a.eps}
    \FigureFile(86mm,86mm){figure5b.eps}
  \end{center}
  \caption{The structure of the accretion flows (5a; left panel) and emergent spectrum (5b; right panel) for $(\alpha, \dot{m}_{\rm tot}, \dot{m}_{\rm corona}/\dot{m}_{\rm disk})$=(1, 1, 99) in the hybrid model.
The thin disk is truncated at $r_{\rm tr}=10 r_{\rm S}$.
Left panel (5a): Same as figure \ref{fig:corona}a but for the hybrid model.
Right panel (5a): Emergent spectrum (solid line) and soft photon spectrum from the thin disk (dashed line).
}
 \label{fig:hybrid}
\end{figure*}

We study the hybrid case, in which a thin disk covered with disk coronae is truncated at $r=r_{\rm tr}$ and transformed to the ADAF .
The structure of the ADAF/corona is shown in figure \ref{fig:hybrid}a for $(\alpha, \dot{m}_{\rm tot}, \dot{m}_{\rm corona}/\dot{m}_{\rm disk}, r_{\rm tr}/r_{\rm S})$=(1, 1, 99, 10).
We can see that the radial gradient discontinuously decreases inside $r_{\rm tr}$ where the underneath thin disk does not exist.
Because of the differences in the radial distribution of the electron temperature (and thus the $y$-parameter), the outer corona and the inner ADAF produce hard X-ray with different power-law indices.
The spectrum has a relatively flat power-law component ($\Gamma_{\rm s} \sim 2$) in the intermediate energy range, $1-10$ keV, generated by the outer corona with small $y<1$, and a hard power-law component ($\Gamma_{\rm h} < 2$) in the high energy range, $>10$ keV, with cut-off at a few hundreds keV, by the inner ADAF with $y \gtrsim 1$ (figure \ref{fig:hybrid}b).
Unlike the inner hot flow model, the hard X-ray component can be dominant, as long as the accretion rate of the corona is sufficiently larger than that of the thin disk (typically $\dot{m}_{\rm corona}/\dot{m}_{\rm disk} \gtsim 100)$, since the thin disk emits less soft photons than that in the inner hot flow model with the same total accretion rate.

\begin{figure*}
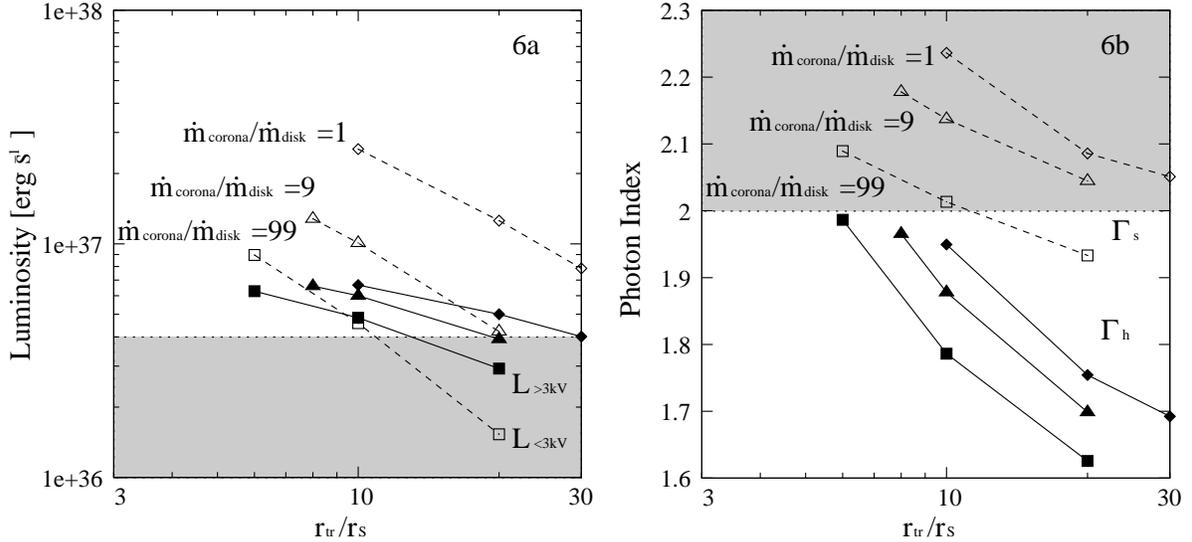

  \begin{center}
    \FigureFile(80mm,80mm){figure6a.eps}
    \FigureFile(77mm,77mm){figure6b.eps}
  \end{center}
  \caption{The spectral features of the hybrid model for $\alpha=1$ and $\dot{m}_{\rm tot}=1$ as functions of the truncation radius, $r_{\rm tr}$. 
The ratio of the accretion rates are $\dot{m}_{\rm corona}/\dot{m}_{\rm disk}=1$ (diamonds), $\dot{m}_{\rm corona}/\dot{m}_{\rm disk}=9$ (triangles), and $\dot{m}_{\rm corona}/\dot{m}_{\rm disk}=99$ (squares).
Left panel (\ref{fig:hybrid2p}a): The luminosity $L_{\rm >3keV}$ and $L_{\rm <3keV}$, is plotted by the filled and open symbols, respectively.
The hard X-ray luminosity should be above the shaded region for the low/hard state spectrum.
Right panel (\ref{fig:hybrid2p}b): Photon index of the softer one of the power-law components, $\Gamma_{\rm s}$ (evaluated by the two points of 3keV and 10keV), and that of the harder one, $\Gamma_{\rm h}$(evaluated by the points of 30keV and 100keV).
The spectrum satisfies the criterion if $\Gamma_{\rm h}$ is below the shaded region.
}
 \label{fig:hybrid2p}
\end{figure*}

We show the features of the emergent spectra for various parameters in figure \ref{fig:hybrid2p}.
We cannot find the advection-dominated solutions for small values of $r_{\rm tr}$ with $\dot{m}_{\rm tot}=1$ because of efficient Compton cooling.
For $r_{\rm tr} \gtsim 6 r_{\rm S}$ we can see the trend that the larger $r_{\rm tr}$ is, the smaller becomes the photon indices of the power-law components and the larger becomes the difference between the two photon indices ($\Gamma_{\rm s}-\Gamma_{\rm h}$).
This means that the break between the two power-law components is more prominent, when the thin disk is truncated at a larger radius.

We see that the soft X-ray luminosity steeply decreases as $r_{\rm tr}$ increases, while the hard X-ray luminosity only gradually decreases.
Thus the hard X-ray component can dominate over the soft one for large $r_{\rm tr} (\gtsim 10 r_{\rm S}$ for $\dot{m}_{\rm corona}/\dot{m}_{\rm disk}=99)$.
This is because the reprocessed radiation becomes weaker when the disk is truncated at a larger radus.
For a very large $r_{\rm tr} (\gg 10 r_{\rm S}$ for $\dot{m}_{\rm corona}/\dot{m}_{\rm disk}=99)$, on the other hand, the absolute value of the hard X-ray luminosity becomes less than that of the criterion.
The ratio of the luminosity $L_{\rm >3keV}/L_{\rm <3keV}$ decreases as the ratio of the accretion rate $\dot{m}_{\rm corona}/\dot{m}_{\rm disk}$ decreases, and for small $\dot{m}_{\rm corona}/\dot{m}_{\rm disk}<10$, it is hard to obtain the results satisfying the criteria for the low/hard state.
For a small total mass accretion rate $\dot{m}_{\rm tot}$, the hard X-ray luminosity is significantly reduced as we saw in the disk corona model (section \ref{sec:corona}).
Finally, we find that the resultant spectra of the hybrid model agree well with the observational results only for $\dot{m}_{\rm tot}=1$, $\dot{m}_{\rm corona}/\dot{m}_{\rm disk} \gtsim 10$, and $r_{\rm tr} \sim 10-20 r_{\rm S}$, though the typical electron temperature is 200-300 keV, which is slightly larger than that estimated by the X-ray observation, $\sim 100$ keV.

\section{Discussions}
\label{sec:discussions}

\subsection{General Discussion}

\subsubsection{Brief Summary}

We calculated the expected spectra from the black hole accretion flows with three different flow geometries.
We find that only the hybrid model can account for the observed spectra in the low/hard state for the following reasons.
(i) It can reproduce the two power-law components from the two separate hot regions having different thermal properties: the outer corona and the inner ADAF.
The former produces the softer power-law component with $\Gamma_{\rm s} \sim 2$, while the latter generates the harder one with $\Gamma_{h}<2$.
(ii) The spectrum is dominated by hard X-ray radiation because of less soft photons originating from the outer thin disk impinging the two hot zone (i.e., we find less dissipated energy in the thin disk compared to the inner hot flow case and less reprocessed radiation compared to the disk corona case).
When the number of the soft photons is small, the soft thermal emission becomes weak and the Comptonization in the hot zone becomes very efficient since $y$ is moderately large ($y \gtrsim 1$).

\subsubsection{On the $\alpha$ Value}

When the optical depth is much smaller than unity, for a moderately large $y \sim 1$, the electron temperature should be very high, $k T_{\rm e} \gtsim m_{\rm e} c^2$.
However, the resulting spectrum does not show a simple power-law component with such a high electron temperature (see figure \ref{fig:corona}b).

Therefore the spectrum can have a strong power-law component only for a moderately large optical depth, $\tau_H \gtsim 1$.
Since the optical depth of the hot advection-dominated flow is given by $\tau_H \sim  \dot{m} / \alpha$, the smaller $\alpha$ is, the larger becomes the optical depth for a given accretion rate.
However the solution of advection-dominated flow exists only when the mass accretion rate is smaller than critical value, $\dot{m}_{\rm crit} \propto \alpha^2$ (Narayan et al. 1995; Abramowicz et al. 1995).
Hence the maximum optical depth is larger for larger $\alpha$, $\tau_H \lesssim \alpha$, and large $\alpha (\sim 1)$ is necessary for a moderately large optical depth of the hot flow, $\tau_H \sim 1$, to be realized (Zdziarski 1998).

Global three-dimensional MHD simulations, on the other hand, showed that the effective value of the viscosity parameter is very small, $\alpha=0.02-0.2$, in the non-radiative ADAF or RIAF (Hawley et al. 2001; Machida \& Matsumoto 2003).
However the ADAF/corona which we consider is radiative and interacts with the thin disk.
Local radiative MHD simulation shows magnetically dominated corona forming above the thin disk (Miller \& Stone 2000; Hirose et al. 2006).
Moreover, when the hot flow collapses in the vertical direction by radiative cooling, we could expect a magnetically supported accretion flow (Machida et al. 2006; Oda et al. 2009).
In the flow supported by the magnetic fields, the effective $\alpha$ could be large, $\alpha \sim 1$, though our model does not take into account the magnetic processes explicitly, and the $\alpha$ prescription might not be relevant in such a low $\beta (<1)$ region.

\subsubsection{Heating of ADAF/corona by Magnetic Fields}

Since the hard X-ray photons are emitted from the ADAF/corona regions where the advection fraction is almost unity, the radiative efficiency of the flow we consider is very small, $L/\dot{M}_{\rm tot} c^2 \sim 0.01$.
This conclusion would be changed, if there are direct heating of electrons by magnetic processes, in addition to the electron heating by the Coulomb collisions with protons.
Moreover since the heating by magnetic fields is non-local process, the energy dissipated in some region can be transported to other distant region via magnetic fields.

In the inner hot flow case, if a significant fraction of energy released in the outer thin disk is dissipated in the inner ADAF region, the radiative efficiency of the thin disk can be sufficiently small $(\ll 1)$, and that of the ADAF becomes large $(<1)$.
As a result, the hard X-ray luminosity from the ADAF can exceed the soft X-ray luminosity from the thin disk.
Dove et al. (1997) and Poutanen et al. (1997) considered similar situation and claimed that inner hot flow model can basically explain the low/hard state spectra. 

In the disk corona case, on the other hand, the conclusion that soft X-ray luminosity is always dominant is hardly affected by the heating process of the corona.
Even if all of the accretion energy of the disk and corona is released by the hard X-ray radiation in the corona, half of the hard X-rays is reprocessed by the thin disk.
Consequently the hard X-ray luminosity cannot be larger than the soft X-ray luminosity (Dove et al. 1997; Cao 2009).

\subsubsection{Mass Loss from the ADAF/corona}

\begin{figure}
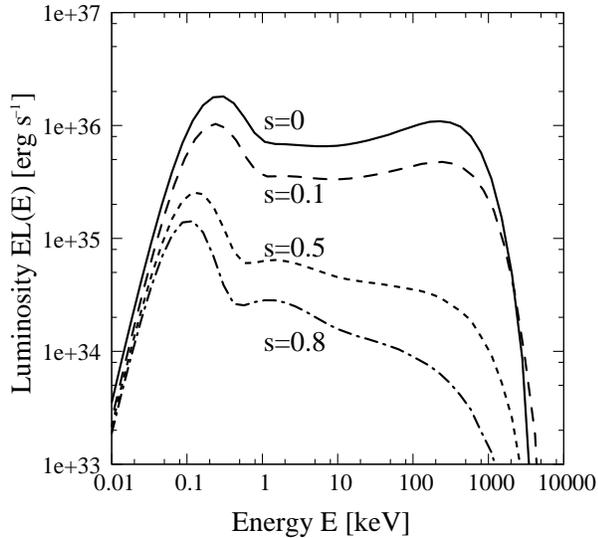

  \begin{center}
    \FigureFile(80mm,80mm){figure7.eps}
  \end{center}
  \caption{The emergent spectrum of the hybrid model for $(\alpha, \dot{m}_{\rm corona}/\dot{m}_{\rm disk}, r_{\rm tr}/r_{\rm S})$=(1, 99, 10).
The total mass accretion rate is assumed to be $\dot{m}_{\rm tot}=\dot{m}_{\rm tot, out} (r/r_{\rm out})^s$ where $\dot{m}_{\rm tot, out}=1$ and $s$=0(solid line), 0.1(long-dashed line), 0.5(short-dashed line), and 0.8(dash-dotted line).}
 \label{fig:massloss}
\end{figure}

So far we assumed no mass loss, thus setting $\dot{m}_{\rm tot}$=const in the radial direction.
We next calculate the effect of the mass loss from the ADAF/corona by using radially dependent mass accretion rate in the case of the hybrid model.
For simplicity, we assumed that the total accretion rate decreases with decreasing radius, $\dot{m}_{\rm tot} = \dot{m}_{\rm tot, out} (r/r_{\rm out})^s$ with $s$ being a constant, while $\dot{m}_{\rm corona}/\dot{m}_{\rm disk}$ is kept constant.
Here $\dot{m}_{\rm tot, out}$ is the total mass accretion rate at the outer boundary, $r=r_{\rm out} (=100 r_{\rm S})$.
In figure \ref{fig:massloss} we show the results of the emergent spectra for various values of $s$.
We find that as $s$ increases the total luminosity decreases and the power-law component becomes softer and weaker compared with the thermal component.
Thus the effect of the mass loss only weakens the hard X-ray component.
If only the accretion rate of the corona decreases with decreasing radius while that of the thin disk is constant ($\dot{m}_{\rm corona}/\dot{m}_{\rm disk}$ decreases), we can expect that the hard X-ray component becomes even weaker than that in the above case. 
Therefore the effect of the mass loss results in an unfavorable consequence for the low/hard state.

\subsection{Observational Implications}

\subsubsection{Photon Index and Truncation Radius}

Zdziarski et al. (1999) found the correlation between the indices of the power-law component and the reflection fractions, $R$, in the accreting systems.
They explained this correlation by changing the inner radius of the outer thin disk surrounding an inner hot flow.
Let us confirm their result by our hybrid model.

\begin{figure}
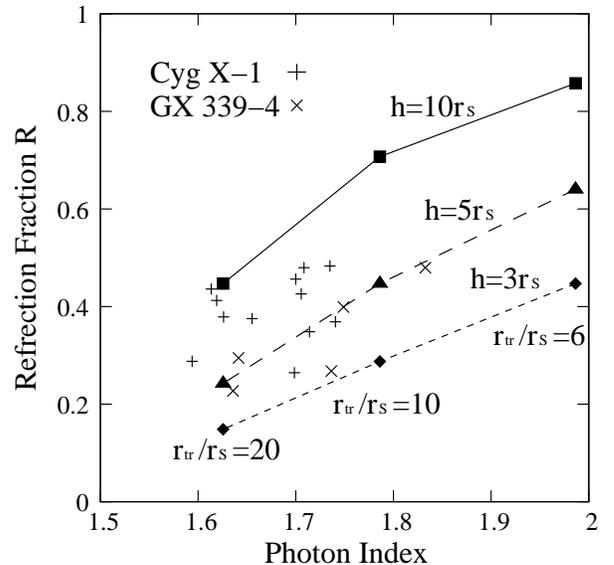

  \begin{center}
    \FigureFile(80mm,80mm){figure8.eps}
  \end{center}
  \caption{Correlation between the photon index $\Gamma_{\rm h}$ and reflection fraction $R$ in the hybrid model for $(\alpha,\dot{m}_{\rm tot},\dot{m}_{\rm corona}/\dot{m}_{\rm disk})$=(1, 1, 99).
In order to calculate $R$, we assume the height of the hard X-ray source, $h=10 r_{\rm S}$(squares), $h=5 r_{\rm S}$(triangles), and $h=3 r_{\rm S}$(diamonds).
The data points are taken from Zdziarski et al. (1999).}
 \label{fig:GR}
\end{figure}

Since we do not include the reflection features in the spectra and it is difficult to estimate the reflection fractions in the complicated accretion structures as are envisaged by our models, in order to calculate the reflection fraction for a given $r_{\rm tr}$, we thus consider a simple model.
Let us consider a point source of hard X-rays located above the central black hole, $(r,z)=(0,h)$, where $h$ is the heigh of the hard X-ray source.
We calculate the solid angle, $\Omega$, subtended by the thin disk which is truncated at $r_{\rm tr}$ and which extends to infinity at the outer part.
Then we obtain the reflection fraction, $R \equiv \Omega/2 \pi$.
In figure \ref{fig:GR} we show the relationship between $R$ and $\Gamma_{\rm h}$ calculate based on our model, overploting the observed values of Cyg X-1 and GX 339-4 taken from Zdziarski et al. (1999).
We find that our model reasonably agrees with the observed correlations, especially for $h=5 r_{\rm S}$, though the data are quite dispersed.
Note that even if the disk is truncated at the radius far from the black hole, significant fraction of the hard X-rays originates from the inner ADAF region.
Hence we expect that the typical position of the hard X-ray does not change so significantly, even when $r_{\rm tr}$ increases.

\subsubsection{Two Power-law Components}

We can reproduce the two power-law components, which are suggested by the observations.
The softer power-law originates from the outer disk corona, where $y$ is relatively small, while the harder one originates from the inner ADAF.
The photon indices of the two power-law components obtained in our model are $(\Gamma_{\rm s}, \Gamma_{\rm h}) \simeq (2.0, 1.8)$ for $(\alpha, \dot{m}_{\rm tot}, \dot{m}_{\rm corona}/\dot{m}_{\rm disk}, r_{\rm tr}/r_{\rm S})$=(1, 1, 99, 10), which roughly agree with the observed values of Cyg X-1, $(\Gamma_{\rm s}, \Gamma_{\rm h}) \sim (1.9, 1.7)$, and GRO J1655-40, $(\Gamma_{\rm s}, \Gamma_{\rm h}) \sim (1.8, 1.6)$ (Takahashi et al. 2008; Makishima et al. 2008).

Note that the two power-law components are produced given that the thin disk is truncated at some radii and that the ADAF/corona are interacting with the soft photons from the thin disk.
In other words the appearance of the two power-law components is a good indication of the disk truncation.
We see that the spectral features depend on $r_{\rm tr}$ (see figure \ref{fig:hybrid2p}), so the spectral evolution of the power-law component provides information regarding the evolution of the truncation radius; the larger $r_{\rm tr}$ is, the larger become the difference between the photon indices, $\Gamma_{\rm s}-\Gamma_{\rm h}$.

Note that there is another explanation of the origin of the two power-law components.
It is well known that the spectrum varies on short time scales.
It is possible that two power-law components could be artificially generated because of the time average in the long observational time.
If one tries to explains also time lags, then one would have to conclude that either the local spectrum is pivoting due to evolution of the geometry of the energy dissipation region (Poutanen \& Fabian 1999), or the spectrum changing because at different distances from the black hole the spectrum of the hot flow is different and it is harder closer in (Kotov et al. 2001).

\subsubsection{Soft Thermal Component}

We evaluate the soft thermal component from the thin disk, including the reprocessed radiation.
It should be noted that the spectrum of the soft photons in our model is significantly different from the disk-blackbody spectrum with an accretion rate of $\dot{M}_{\rm disk}$.
Rather it exhibits a high effective temperature (or large spectral hardening factor, $\kappa$), because the contribution of the reprocessed radiation of the hard X-ray from the ADAF/corona is large (typically $\gtsim 50 \%$ of the soft photons).
Hence, the simple estimation of the inner disk radius by the disk-blackbody model would underestimate the actual disk radius (Gierli\'{n}ski et al. 2008).

\subsubsection{Spectral Transitions}

We find that the spectra show very different spectral features for various ratios of the accretion rate of the corona to that of the thin disk, even when the total accretion rate is kept constant.
The larger $\dot{m}_{\rm corona}/\dot{m}_{\rm disk}$ is, the larger becomes the fraction of hard X-rays (see figure \ref{fig:hybrid2p}a).
This implies that a change in the ratio of the accretion rate may trigger a transition of the spectral state, even when the total mass accretion rate is nearly kept constant.
Note that our model produce a thermal cut-off in the power-law component around $\sim 100$ keV.
Thus the non-thermal electrons would be necessary for the spectrum in the high/soft state, and also in the low/hard state if the observed radiation in the MeV band comes from the accretion flow (Li et al. 1996; Poutanen \& Coppi 1998; Gierlinski et al. 1999; McConnell et al. 2002; Yuan et al. 2003; Inoue et al. 2008).

\subsection{Further Developments of Our Model}

The electron temperature expected by our model (typically 200-300keV) is somewhat higher than that estimated by the cut-off energy of the hard X-ray spectrum ($\sim$100keV).
This is because, even though the viscosity parameter is large, $\alpha \sim 1$, and the mass accretion rate is its maximum value $\dot{m}_{\rm tot}$, the optical depth of the ADAF is still somewhat smaller than unity, resulting in high electron temperature in order to keep large Compton $y$-parameter, $y \sim 1$ (Zdziarski 1998).
The optical depth of the electron-positron pairs could contribute to lower the electron temperature, since the compactness parameter around the black hole in the low/hard state is expected to be large,  
\begin{eqnarray}
l \equiv \frac{L \sigma_T}{R m_{\rm e} c^3} \sim 10 \left( \frac{L}{0.01 L_{\rm Edd}} \right) \left( \frac{R}{10 r_{\rm S}} \right)^{-1},
\end{eqnarray}
where we set the typical size of the emitting region, $R = 10 r_{\rm S}$.
The newly found solutions, luminous hot accretion flows, might also be appropriate, since they have large optical depth and low electron temperatures, compared with ADAFs for a given $y$ (Yuan 2003; Yuan \& Zdziarski 2004), though they might be thermally unstable.
Moreover thermal synchrotron emission would play a role as a cooling mechanism of the electrons, in addition to a role as a source of the soft photons.

We find that the hybrid model with the parameters used in this paper can marginally agree with the relatively faint low/hard state ($L/L_{\rm Edd} \sim 0.01$).
However the hard X-ray luminosity obtained by our model is still fainter than the relatively luminous low/hard state, $L/L_{\rm Edd} \gtsim 0.1$.
In order to explain this state, we could consider other accretion solutions of the ADAF/corona: e.g., luminous hot accretion flow (Yuan \& Zdziarski 2004) or magnetically dominated accretion flow (Oda et al. 2009).

In our model we simply give the transition radius and mass accretion rate as free parameters.
However these values seem to depend on the Compton cooling rate in the disk evaporation model (Liu et al. 2002).
Hence we have to solve the whole of the structure, taking in to account the gas evaporation/condensation process and radiative transfer self-consistently.
In addition we used the self-similar solution for ADAF/corona, which would be invalid near the black holes where boundary conditions become essential (Nakamura et al. 1996; Manmoto et al. 1997; Narayan et al. 1997).
For further studies, we should include general relativistic effect into both accretion structure and radiative transfer (doppler beaming by the gas motion and bending of the photon trajectory), in order to include the radiative contribution from the plunging region where the bulk Comptonization would also be important.

\section{Conclusions}
\label{sec:conclusions}

We calculate the emergent spectra from three accretion geometries: (a) inner hot flow model, (b) disk corona model, and (c) hybrid model.
We find following results.

1. Only the hybrid model satisfies all the basic features of the low/hard state for $\dot{m}_{\rm corona} \sim 1$, $\dot{m}_{\rm disk} \ll \dot{m}_{\rm corona}$, and $r_{\rm tr} \sim 10 r_{\rm S}$.
Other models predict weak hard X-ray emission and at most one power-law component.

2. Two power-law components simply imply that the thin disk is truncated and there are two regions with different thermal properties: the outer corona produces softer component dominating at $<10$ keV and the inner ADAF generates a harder component. 

3. The hybrid model can also explain the observed $\Gamma-R$ relation.

\bigskip

The authors are grateful to K. Ohsuga for useful discussions and comments.
The authors also thank to anonymous referee for valuable suggestions.
This work is supported in part by the Grant-in-Aid for the global COE programs on "The Next Generation of Physics, Spun from Diversity and Emergence" from the Ministry of Education, Culture, Sports, Science and Technology (MEXT) of Japan, by the Grant-in-Aid of MEXT (19340044, SM), and by Department of Astronomy, Kyoto University.

\end{document}